\documentclass[12pt]{spieman}  
\usepackage{amsmath,amsfonts,amssymb}
\usepackage{graphicx}
\usepackage{setspace}
\usepackage{tocloft}
\usepackage{subcaption}

\title{Multiple-Amplifier Sensing Charged-Coupled Device: Model and improvement of the Node Removal Efficiency}

\author[a,b,*]{Blas J. Irigoyen~Gimenez}
\author[a,b]{Miqueas E. Gamero}
\author[b]{Claudio R. Chavez~Blanco}
\author[a,c]{Agustin J. Lapi}
\author[a,c]{Fernando Chierchie}
\author[b,d]{Guillermo Fernandez~Moroni}
\author[b]{Juan Estrada}
\author[b]{Javier Tiffenberg}
\author[b,d,e]{Alex Drlica-Wagner}
\affil[a]{Universidad Nacional del Sur, Bahia Blanca, Argentina.}
\affil[b]{Fermi National Accelerator Laboratory, P.O.\ Box 500, Batavia, IL 60510, USA.}
\affil[c]{Insituto de Investigaciones en Ing. Eléctrica "Alfredo Desages" (IIIE) CONICET, Argentina.}
\affil[d]{Department of Astronomy and Astrophysics, University of Chicago, 5640 South Ellis Avenue, Chicago, IL, 60637, USA.}
\affil[e]{Kavli Institute for Cosmological Physics, University of Chicago, Chicago, IL 60637, USA.}

\cftpagenumbersoff{figure}
\cftpagenumbersoff{table} 

\begin{document} 
\maketitle
\begin{abstract}

The Multiple Amplifier Sensing Charge-Coupled Device (MAS-CCD) has emerged as a promising technology for astronomical observation, quantum imaging, and low-energy particle detection due to its ability to reduce the readout noise without increasing the readout time as in its predecessor, the Skipper-CCD, by reading out the same charge packet through multiple inline amplifiers. Previous works identified a new parameter in this sensor, called the Node Removal Inefficiency (NRI), related to inefficiencies in charge transfer and residual charge removal from the output gates after readout. These inefficiencies can lead to distortions in the measured signals similar to those produced by the charge transfer inefficiencies in standard CCDs. This work introduces more details in the mathematical description of the NRI mechanism and provides techniques to quantify its magnitude from the measured data. It also proposes a new operation strategy that significantly reduces its effect with minimal alterations of the timing sequences or voltage settings for the other components of the sensor. The proposed technique is corroborated by experimental results on a sixteen-amplifier MAS-CCD. At the same time, the experimental data demonstrate that this approach minimizes the NRI effect to levels comparable to other sources of distortion the charge transfer inefficiency in scientific devices.

\end{abstract}

\keywords{Charge coupled device, Astronomy instrumentation, Single photon counting imager,Non-destructive readout sensor, Single electron resolution imager, MAS-CCD, Multiple distributed amplifiers,.}

{\noindent \footnotesize\textbf{*}Blas J. Irigoyen~Gimenez,  \linkable{blasiri16@gmail.com} }

\begin{spacing}{2}  

\section{Introduction}
\label{sect:intro}

\begin{figure*}[t!]
    \centering
    \includegraphics[width=1\textwidth]{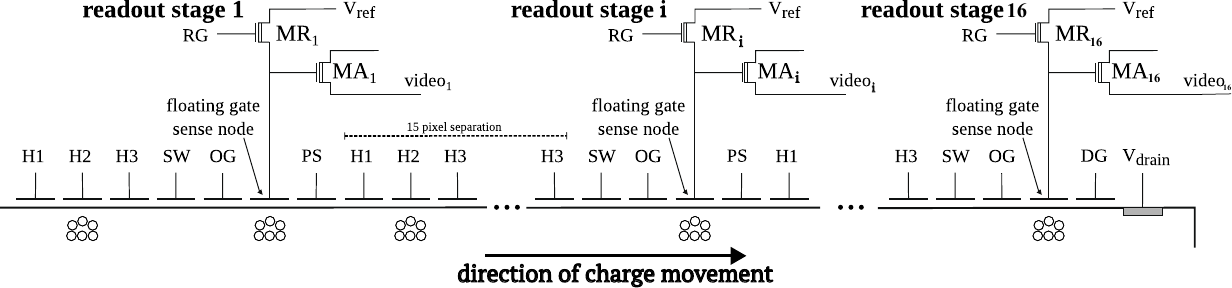}
    
    \caption{Architecture of the sixteen inline amplifiers at the end of the serial register of the MAS-CCD. The charge packets flow through the sensor's channel underneath the gates in the direction denoted by the arrow as the effect of the change of the potential's value in the channel generated by the clocks applied to the gates.}
    \label{fig:architecture}
\end{figure*}

The Multiple-Amplifier Sensing Charge Coupled Device (MAS-CCD)\cite{holland_2023} offers the advantage over the Skipper-CCD \cite{Tiffenberg:2017aac} in reducing readout noise without significantly increasing readout time \cite{MASCCD8_2024,MASCCD16_2024} by reading out the exact charge packet through multiple inline amplifiers. This architecture is shown in Fig.  \ref{fig:architecture}. The charge of the pixels is moved through the serial register, and the different amplifiers measure its value. When the information is combined, the readout noise is reduced by the square root of the number of amplifiers. At the same time, each output stage can perform multiple non-destructive samples thanks to its floating gate amplifier (FGA) sense node, which further reduces the readout noise as the square root of the samples per pixel. The sensor was identified as a candidate technology for constructing the next generation of scientific spectroscopic experiments.\cite{MASCCD16_2024,lin2024,RauscherNASA2022}

Figure \ref{fig:architecture} examples the movement of the train of charge packets during readout. To enable the charge transfer between consecutive amplifiers, pixel-separation gates (PS) facilitate the removal of charge from the sense node to the following chain of horizontal gates (H1, H2, H3), which provide the three-clock sequence to move the charge between consecutive stages (Fig.\ref{fig:architecture}). The dump gate (DG) and drain (V$_{\text{drain}}$) contacts are used to remove the charge from the channel after its measurement in the last amplifier. In each stage, MR$_{\text{i}}$ is the reset gate transistor which is turned on to set the reference voltage $V_{\text{ref}}$ on the sense node before measuring the charge of a pixel. The output transistor (MA$_{\text{i}}$) converts the charge into a voltage in a source follower configuration polarized with V$_{\text{dd}}$ and 20k$\Omega$ resistor. For a detailed description of the MAS architecture, please refer to \cite{MASCCD8_2024}. 

A new performance parameter called Node Removal Efficiency (NRE) has been identified exclusively for MAS-CCD to account for the inefficiencies in removing the charge from the output stages. During this process, some of the charges can be left behind in the sense nodes, which are then collected by the next pixel in the line. Independent measurements in \cite{MASCCD16_2024} and \cite{lin2024} have shown the distortion associated with the mechanism and a detailed analytical model is presented in \cite{MASCCD16_2024}. The findings of the inefficiency mechanism in previous work were used to develop a new operation scheme for the sensor, which involves the use of an extra clock signal on the sensor to improve the NRE without the need to change the time sequence or voltage operation of the rest of the gates in the sensor. An explanation of the proposed technique and the experimental results showing the improvements in the NRE are shown in this article.

\section{Node removal inefficiency and compensation technique}

\begin{figure*}[h!]
    \centering
    \includegraphics[width=0.82\textwidth]{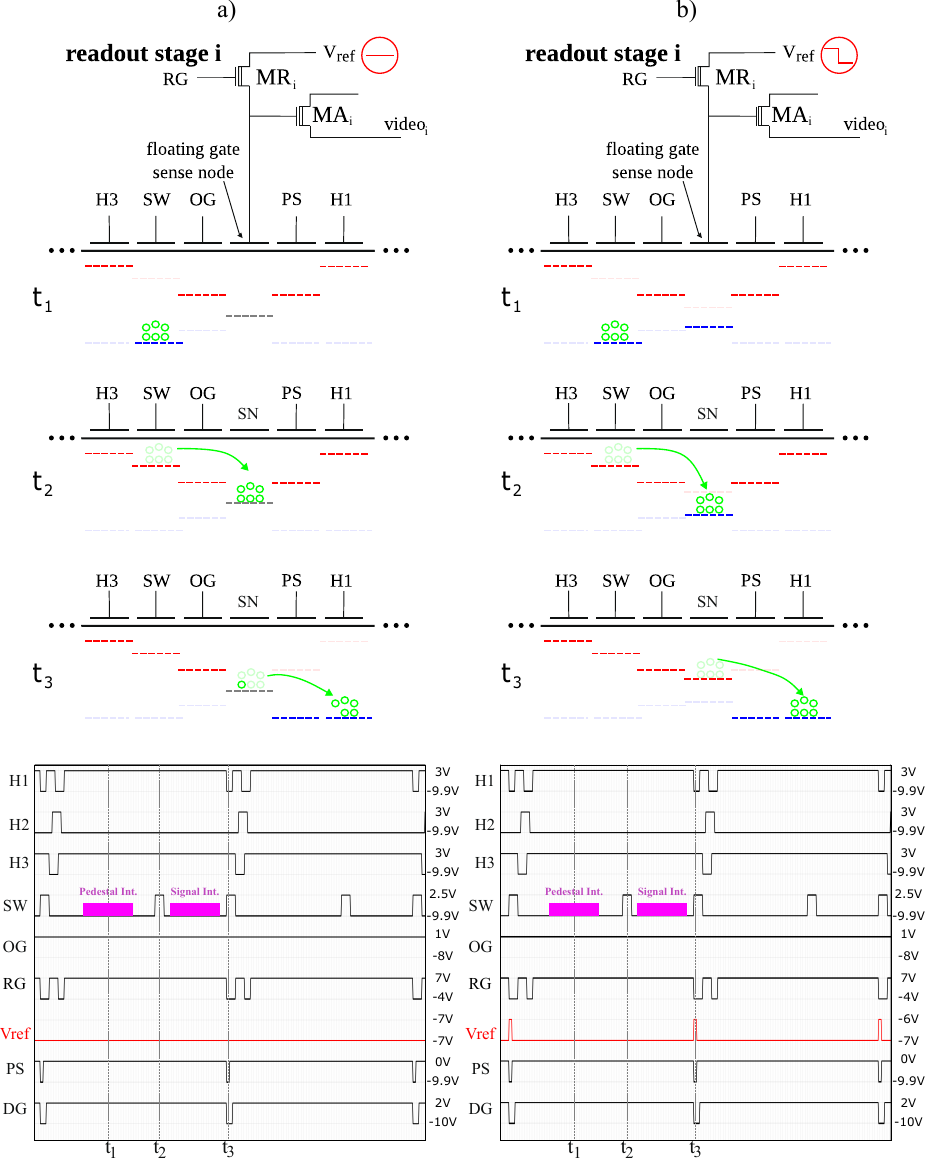}
    
    \caption{Scheme of the readout process in a single intermediate output stage with the corresponding voltage levels at three times: $t_1$, when the pedestal level is measured, $t_2$, when the charge is dumped to the SN to measure the signal level, and $t_3$, when the charge packet is extracted from the sense node after being measured, a) without the technique proposed: the fraction of charge left in the sense node (SN) after the readout due to the problem of the NRE is illustrated, and b) with the technique proposed: clocking the reference voltage $V_{\text{ref}}$ to be able to make it possible to remove all the charge from the sense node after it is readout. The dashed line shows a qualitative potential value in the channel generated by the clocks applied to the gates to move the charge through the different structures. The approximate integration windows for the signal and pedestal levels are shown for both cases.}

    \label{fig:clk_schm}
\end{figure*}

Due to the NRE, a fraction of the measured charge by one amplifier may remain in the sense node. Figure \ref{fig:clk_schm}a) illustrates this mechanism at different times for an intermediate amplifier of the MAS-CCD. The diagram at the bottom shows the clocking sequence involving the charge transfer in the output stages. At time $t_1$ the charge packet is under the SW gate while the readout electronics measure the reference value (pedestal level) on the floating SN. The voltage in the SN was previously set to V$_{ref}$ when the RG signal was set low.

Later, at $t_2$, the SW gate is set to a high level to dump the charge packet to the sense node, then set to the low level again, and then the signal level measurement is performed to complete the pixel calculation through the Dual Slope Integrator (DSI). After that, this measured charge packet must be removed from the SN and transferred under the next $H_2$ clock. To accomplish this, at $t_3$, the PS clock is set to a low level together with the $H_1$ clock. During this process, some of the charges could be left behind due to limited potential gradient while transferring large signal packets, limited transfer periods from the fast operation of the sensor, or traps or potential distortions in the channel. Figure \ref{fig:clk_schm}a) shows that some of the carriers are still under SN at $t_3$. 

The remaining charge will be part of the reference (pedestal) level of the following pixel for the DSI calculation, and therefore it will not be measured by amplifier in stage $i$. However, it will be added up to the charge packet and will be transferred all together to the next amplifier in the next stage (stage $i+1$). Thus, the net results appear at the time the pixel is measured by the subsequent amplifiers. For this reason, as we will see in the experimental results, the NRE does not exist in the output data from the first amplifier of the chain. This is a strong confirmation of the mechanism producing the NRE and is useful to identify and isolate NRE from other possible transfer issues. Each amplifier contributes to the NRE, so its effect is proportional to the number of amplifiers in the chain.

In a more rigorous way, following the description in \cite{MASCCD16_2024}, the value of the charge packet $q_i[n]$ (time instant $n$) measured in (at the sense node of) the amplifier $i$ as a function of the charge in the same charge packet but measured by the previous amplifier $q_{i-1}[n - N_{p}]$ where $N_{p}$ is the number of intermediate serial register pixels between amplifiers, and the previous charge packet $q_{i}[n-1]$ at the previous time $(n-1)$ at the same amplifier $i$ can be used to get a recursive equation to emulate the NRE problem:

\begin{align}
    q_i[n]&=(1-\epsilon)\left(q_{i-1}[n-N_{p}-1]+\frac{\epsilon}{(1-\epsilon)} q_{i}[n-1]\right) \nonumber \\
    &=(1-\epsilon)q_{i-1}[n-N_{p}-1]+\epsilon q_{i}[n-1]
    \label{eq:Recursive_qj}
\end{align}
where $(1-\epsilon)$ is  the NRE and $0\leq \epsilon \leq 1$ is the symbol used for to account for the NRI.

The first line in Eq.(\ref{eq:Recursive_qj}) represents the efficiency $(1-\epsilon)$ of transferring the charge packet that was in the previous amplifier $(i-1)$ sense node to the next amplifier $i$, with $i\geq 2$. This charge packet has two components: the first part that was effectively measured by the previous amplifier $q_{i-1}[n-N_{p}-1]$ and secondly the fraction of the charge that was present in the SN of the amplifier $(i-1)$ but was not measured because it was part of the reference voltage (pedestal level). This fraction of charge was lost from the charge packet that was previously measured in the sense node of amplifier $i$ (i.e. $q_{i}[n-1]$) and can be estimated from it as $\frac{\epsilon}{(1-\epsilon)} q_{i}[n-1]$, where $q_{i}[n-1]/(1-\epsilon)$ estimates the charge packet when it was in the previous amplifier and that times the inefficiency $\epsilon$ estimates the fraction of charge in the second term.

Equation \ref{eq:Recursive_qj} can be used to model the charge spilled over in the overscan pixels by the last column of the active region. We assume for the first amplifier, that the first pixel in the overscan column occurs at time $n_{o,1}$ and that the last pixel of the active region column has a charge of $Q$. As we explained before, the measurement of the first amplifier does not see the NRI effect, so $q_1[n_{o,1}]=0$. Using Eq. \ref{eq:Recursive_qj}, the amount of charge measured by the other amplifiers in the first pixel of the overscan, at index  $n_{o,i}=n_o+(i-1)(N_{p}+1)$, is:
\begin{align*}
   q_2[n_{o,2}] &= (1-\epsilon)q_{1}[n_{o}] + \epsilon q_{2}[n_{o} + N_{p} - 1] = \epsilon Q \\
   q_3[n_{o,3}] &= (1-\epsilon)q_{2}[n_{o} + N_a] + \epsilon q_{3}[n_{o} + 2N_{p} - 1] = (1-\epsilon)\epsilon Q + \epsilon Q = \epsilon Q(1 + (1 - \epsilon)) \\
   &\vdots\\
   q_i[n_{o,i}] &= (1-\epsilon)q_{i-1}[n_{o} + (i-2)N_{p}] + \epsilon q_{3}[n_{o} + 2N_{p} - 1] = \epsilon Q \sum^{i-1}_{k=0} (1-\epsilon)^k.
\end{align*}

The last line of the equation gives a way to model the expected signal in the first pixel of the overscan due to the NRI. Using geometric series properties, we can condense it to: 
\begin{equation}
    q_i[n_{o,i}] =Q(1-(1-\epsilon)^{i-1}) 
    \label{eq: charge overscan channel i}
\end{equation}
which is an increasing function of the amplifier index. It provides a simple way to estimate the $\epsilon$ based on the measurements of the individual amplifiers. If the information of all the amplifiers are averaged, the first pixel in the overscan will have value
\begin{equation}
    q_{ave}[n_{o,ave}] = \epsilon_{ave}{Q} = \sum^{N_{a}}_{i=1}\frac{q_i[n_{o,i}]}{N_{a}} =\sum^{N_{a}}_{i=1} \frac{Q}{N_{a}}(1-(1-\epsilon)^{i-1})) =\frac{\epsilon{N_a}+(1-\epsilon)^{N_a}-1}{\epsilon{N_a}}  
    \label{eq: eps average}
\end{equation}
where $\epsilon_{ave}$ is the NRI effect measured in the averaged image and can be used to calculate the $\epsilon$ on each amplifier using the measurement in the average image. If we assume that $\epsilon<<1$ in equation \ref{eq: eps average} then 
\begin{equation}
    \epsilon \approx 2\epsilon_{ave}/(N_a-1)
    \label{eq:epsilon from epsilon average}
\end{equation}

\subsection{New clocking scheme}
A new scheme is proposed in this work to explore the Node Removal Efficiency (NRE). It involves using an extra clock signal connected to $V_{ref}$ with a double purpose: its low level is used to set the reference voltage in the SN before dumping the pixel charge into it, and its high level ($V_{ref, high}$) is used to promote the charge drift in the transfer direction. Although the scheme requires a new clock to operate the sensor, it provides flexibility in optimizing its operation. For example, the NRE can be improved by increasing the time interval for its transfer, but as shown in \cite{MASCCD16_2024} its effect could be already appreciated even when reading the sensors at mid-speeds of $70$-kilo pixels per second. Reducing the low value of PS and H clocks is another way to increase the electric field and therefore, the drift for the carrier transport; however, increasing the horizontal swing of the clocks has shown an increase in the spurious charge generated in the serial register \cite{cababie_2022,villalpando_SOAR2024}. Another related aspect is that the polarization of the MA$_{\text{i}}$ amplifier requires a negative voltage of several volts in the SN (gate of the MA$_{\text{i}}$ transistor), imposing a significant negative value for the clocks driving the other gates. Setting even lower voltages to improve the NRE could sometimes be hardware-limited. 

Figure \ref{fig:clk_schm}b) shows the operation used to test the new idea. $V_{\text{ref}}$ is kept at the same low potential during the charge measurement at $t_1$ and $t_2$ but it is transitioned to high level while the PS and H1 is set to low level for the clock to help transporting the charge to the H gates as illustrated at time $t_3$. 

The next subsection shows the experimental results. We focused on two main aspects, showing that the NRE technique can be improved by increasing the potential at the SN, giving quantitative information on the required increase of the $V_{\text{ref}}$ signal, and lately proving that this extra swing does not have an effect in the readout noise of the system. 

\section{Experimental results}
\label{exp_results}

The sensor used for this work consists of $N_{a} = 16$ non-destructive readout stages and possesses a 1024 $\times$ 512 array of 15\,$\mu$m $\times$ 15\,$\mu$m pixels. It was fabricated under the auspices of the U.S. Department Of Energy (DOE) Quantum Information Science initiative. It is a $p$-channel 675$\mu$m-thick MAS-CCDs fabricated on high-resistivity $n$-type silicon (${\sim}\,10$ k$\Omega$\,cm). The sensors were designed at LBNL to be operated as thick fully-depleted devices with high quantum efficiency over a broad wavelength range \cite{Holland:2003} (see \cite{holland_2023} for more details on the fabricated sensors). The sensors were fabricated at Teledyne DALSA Semiconductor, diced at LBNL, and packaged/tested at Fermilab. The output amplifiers are separated by $N_{p} = 15$ intermediate pixels. The sensor is operated at a temperature of 140K and in full depletion mode using a substrate voltage of 70V. To implement the technique presented here, we had to adapt the existing electronic chain to allow having two voltage levels for $V_{\text{ref}}$. 

\begin{figure}[H]
    \centering
    \includegraphics[width=0.4\textwidth]{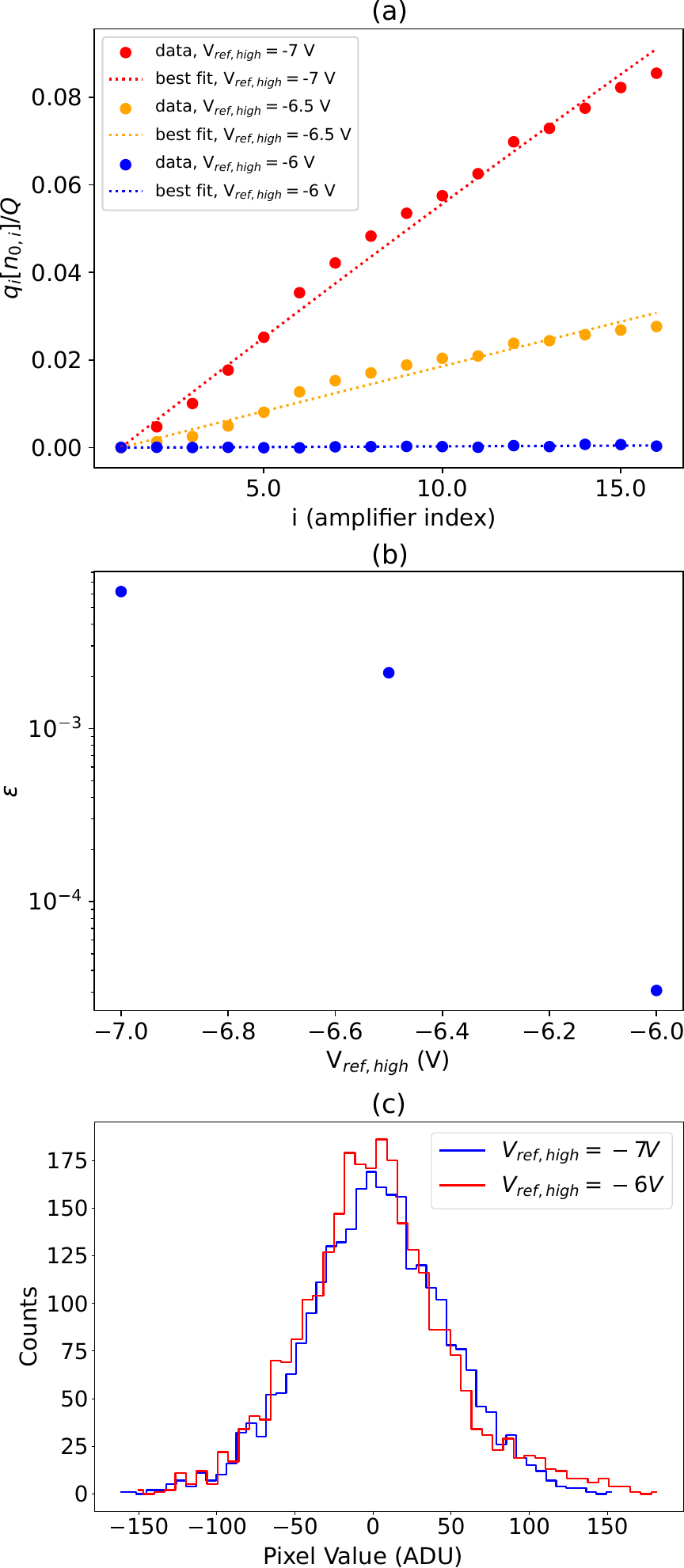}
    \caption{(a) Fraction of the charge that is measured in the first pixel of the overscan as a function of the amplifier index and for different maximum voltages of the $V_{\text{ref}}$ signal. (b) NRI measured in the averaged image. (c) Comparison of the histogram of the pixels in the overscan region for the more extreme condition of the maximum value of the $V_{\text{ref}}$ signal.}
    \label{fig:results}
\end{figure}

The sensor was read at 70-kilo pixels per second with a single sample per amplifier, and the time $t_3$ used for the experimental technique was 500 nanoseconds, as well as the other horizontal transitions to move the charge through the serial register using a three-phase configuration. The same sequencer was used for all the test conditions, except for the $V_{ref, high}$ voltage in the $t_3$ period from Fig. \ref{fig:clk_schm}b). The experimental results are summarized in the plots in Fig. \ref{fig:results}. For this case, three different high voltages $V_{ref, high}$ were used: -7 V (which is the same as the low-level value, so there is no net change in the SN potential), -6.5 V, and -6. The voltages of the remaining gates in the horizontal register are shown in Figure \ref{fig:clk_schm}.

Figure \ref{fig:results} shows the charge observed in the first column of the overscan normalized to the charge of the last column in the active region ($q_i[n_{o,i}]/Q$). Since the overscan pixels are expected to be empty, this quantity tells the proportion of the charge left behind due to the NRI. Ten pixels were used to calculate the mean values of each column. The average signal value in the last column of the active region is 74,000 collected carriers. The fraction of missing carriers is shown as a function of the amplifier index. The red curve is obtained when the V$_{ref}$ signal is at a steady potential at $-7V$ without transitions. Two aspects are interesting to mention here. First, the fraction of the charge left behind is cumulative with the amplifier index as expected. Second, the measurements for the first amplifier show no appreciable NRI, which proves that the problem of the NRI is related to moving the charge out of the SN and not to processes occurring beforehand, such as the horizontal or vertical charge transfer problem. The maximum fraction deposited in the first column in the overscan is larger than 8\%. 
The orange curve in the same figure is obtained when $V_{ref, high}$ is set to $-6.5$V, i.e., half a volt swing. The plot shows a similar trend, but the NRI effect is substantially lower. When the swing is changed to one volt ($V_{ref, high} = -6$V), as shown in the blue curve, the effect of the NRI becomes imperceptible for all the amplifiers which prove to be a powerful tool to mitigate or drastically reduce the inefficiency problem in this kind of sensors. 

Equation \ref{eq: charge overscan channel i} can be used to estimate $\epsilon$ from the presented data. Figure \ref{fig:results}(a) shows the best fits for each case in dotted lines. The estimated parameters are shown in the second column of Table \ref{tab:epsilon}. The $\epsilon$ can also be calculated using the final data product of these types of sensors after averaging the information of all the amplifiers. Equation \ref{eq:epsilon from epsilon average} gives a way to do so as a function of the average fraction of charge in the first column of the overscan. Figure \ref{fig:results} shows the results for each $V_{ref, high}$ considered. The third column of Table \ref{tab:epsilon} shows the numerical values.

The last part of the experimental phase was to estimate if there was any impact on the readout noise performance of the output images while switching the $V_{ref}$ between two values compared to the steady condition. For this purpose, a set of 180 images was taken for $V_{ref, high} = -6$ V and $V_{ref, high} = -7$ V. The variance was calculated on regions of 2400 pixels in the overscan for each image. Figure \ref{fig:results} shows the histogram of pixels for one image from each set, showing no visible difference. The variance is averaged for each dataset and compared. This methodology allows us to compare deviations of the noise as small as $0.1e^-$ between the two datasets, and the measured difference was smaller than that. 

The experimental results show that a small increase in the potential voltage of the sense nodes is an effective way to improve the NRI without changing the readout sequence of the rest of the clocks and without affecting the noise performance. At the same time, the lowest obtained $\epsilon$ value reaches the proposed requirement in \cite{MASCCD16_2024} where the signal distortion caused by the NRI equals the distortion caused by the charge inefficiency transfer (CTI) distortion when the CTI is 1$\times10^{-6}$ on large devices of 2000 columns per quadrant.

\begin{table}[h!]
\centering
\begin{tabular}{|c|c|c|}
\hline
$V_{ref, high}$ [V] & $\epsilon$ (from fit Eq. \ref{eq: charge overscan channel i}) & $\epsilon$ (from Eq. \ref{eq:epsilon from epsilon average}) \\ \hline
-7   & $6.34\times10^{-3}$ & $6.15\times10^{-3}$\\ \hline
-6.5 & $2.08\times10^{-3}$ & $2.09\times10^{-3}$\\ \hline
-6   & $3.24\times10^{-5}$ & $3.24\times10^{-5}$\\ \hline
\end{tabular}
\caption{Measured $\epsilon$ before and after averaging the information from the different amplifiers.}
\label{tab:epsilon}
\end{table}

\section{Conclusions}

A more complete model of the Node Removal Inefficiency mechanism was explained in this article. The model was used then to quantify the effect. A new technique with minimum intrusive changes to the standard readout sequence and operation voltages was presented to mitigate the NRI. The technique's advantages were corroborated by experimental results. The NRI was reduced to the desired levels comparable to other sources of signal distortions from charge transfer inefficiencies in scientific CCDs.

\section* {Acknowledgments}
The fully depleted Skipper-CCD was developed at Lawrence Berkeley National Laboratory, as were the designs described in this work. The CCD development work was supported in part by the Director, Office of Science, of the U.S. Department of Energy under No. DE-AC02-05CH11231.
The multi-amplifier sensing (MAS) CCD was developed as a collaborative endeavor between Lawrence Berkeley National Laboratory and Fermi National Accelerator Laboratory. Funding for the design and fabrication of the MAS device described in this work came from a combination of sources including the DOE Quantum Information Science (QIS) initiative, the DOE Early Career Research Program, and this work was produced by Fermi Forward Discovery Group, LLC under Contract No. 89243024CSC000002 with the U.S. Department of Energy, Office of Science, Office of High Energy Physics. Publisher acknowledges the U.S. Government license to provide public access under the DOE Public Access Plan.
This research has been partially supported by a Detector Research and Development New Initiatives seed grant at Fermilab, a grant from the Heising-Simons Foundation (\#2023-4611), and Javier Tiffenberg's and Guillermo Fernandez Moroni's DOE Early Career research programs.


\bibliography{main}   
\bibliographystyle{spiejour}

\vspace{1ex}
\listoffigures
\listoftables

\end{spacing}
\end{document}